\documentclass[usegraphicx]{aastex702}
\usepackage{amssymb}
\usepackage{bm}
\usepackage{color}
\usepackage{graphicx}

\input epsf

\def\ltsima{$\; \buildrel < \over \sim \;$}
\def\lsim{\lower.5ex\hbox{\ltsima}}
\def\gtsima{$\; \buildrel > \over \sim \;$}
\def\gsim{\lower.5ex\hbox{\gtsima}}

\def\aap{A\&Ap}

\def\apjl{ApJ}
\def\apjs{ApJS}

\def\spose#1{\hbox to 0pt{#1\hss}}
\def\lta{\mathrel{\spose{\lower 3pt\hbox{$\mathchar"218$}}
     \raise 2.0pt\hbox{$\mathchar"13C$}}}
\def\gta{\mathrel{\spose{\lower 3pt\hbox{$\mathchar"218$}}
     \raise 2.0pt\hbox{$\mathchar"13E$}}}

\begin{document}

\title{The heating rate of the Intergalactic Medium by Lyman-$\alpha$ photon scattering}

\author{Avery Meiksin}
\email{E-mail:\ meiksin@ed.ac.uk}
\affiliation{Institute for Astronomy, University of Edinburgh\\
  Royal Observatory of Edinburgh\\
  Edinburgh\ EH9\ 3HJ, UK}
\altaffiliation[Affiliate of\ ]{Scottish Universities Physics Alliance (SUPA)}

\date{\today}

\begin{abstract}
  The strength of the 21-cm Cosmic Dawn radio absorption signature against the Cosmic Microwave Background or against a bright background radio source depends on the temperature of the neutral hydrogen in the intergalactic medium. While x-rays from forming galaxies will likely dominate the heating rate, recent models suggest scenarios in which heating by the scattering of Lyman-$\alpha$ photons sourced by the galaxies contributes non-negligibly as well, and may even dominate. The efficiency of Lyman-$\alpha$ photon heating for a point source at high redshift using a numerical solution to the exact radiative transfer equation is provided here. It is found to be a factor of several smaller than the rate computed in the commonly used diffusion approximation to the radiative transfer equation.
\end{abstract}

\keywords{cosmology:\ theory -- dark ages, reionization, first stars --
intergalactic medium -- radiative transfer -- radio lines:\ general -- scattering
}

\section{Introduction}
\label{sec:intro}

The search for the 21-cm Cosmic Dawn signature of the first major
radiating sources in the Universe has driven the development of a new
generation of radio telescopes, as well as various codes to infer the
implications of any potential signal for galaxy formation. One of the
key ingredients of the simulations is the heating of the intergalactic
gas, as the hydrogen 21-cm absorption strength against the Cosmic
Microwave Background or a bright radio source scales inversely with
the gas kinetic temperature.

The principal heating mechanism of the Intergalactic Medium prior to its reionization is expected to be through x-rays generated by high-mass binary systems
and diffuse emission from gas within galaxies heated by
supernovae. The recoil of Lyman-$\alpha$ photons (and higher
Lyman-series photons at a lesser level) arising from source continuum photons redshifting into a local resonance line provides a secondary heating
source. In some models for young forming galaxies, this contribution
is non-negligible and is included in the design of simulations for predicting the 21-cm Cosmic Dawn signature \citep{2013MNRAS.428.1755C, 2021MNRAS.506.5479R, 2026RASTI...5ag001M, 2026ApJ...996...44R}.

Estimates in the literature for the heating rate by Lyman-$\alpha$
photon recoils are based on approximate solutions to the radiative
transfer equation. The diffusion approximation lends itself to ready
estimates to the heating rate at different redshifts and for different
radiation sources and gas temperatures of the intergalactic gas. While
different approximations to the radiative transfer equation have
solutions that agree far from line center, they differ within the line
core where the scatters dominating heat exchange with the gas occur. A
comparison with an exact (numerical) solution for a point source in an
expanding medium corresponding to typical conditions during Cosmic
Dawn is provided here.

\section{Lyman-$\alpha$ radiative transfer and recoil heating}
\label{sec:RT}

The radiative transfer equation describing the rate of change in the
energy of the radiation field at frequency $\nu$ at position ${\bf r}$
at time $t$ moving (in the laboratory frame) into direction ${\bf \hat
  n}$ per unit area per unit solid angle per unit time per unit
frequency is
\begin{equation}
\frac{D I_\nu({\bf r},t,{\bf \hat n})}{D t}
%=\frac{\partial I_\nu({\bf r},t,{\bf \hat n})}{\partial t}+
%c{\bf \nabla} I_\nu({\bf r},t,{\bf \hat n})
=c\eta_\nu({\bf r},{\bf \hat n},t)-
c\chi_\nu({\bf r},{\bf \hat n},t)I_\nu({\bf r},t,{\bf \hat n}),
\label{eq:RTgen}
\end{equation}
\citep{1978stat.book.....M}, where $I_\nu({\bf r},t,{\bf \hat n})$ is the specific intensity,
$\eta_\nu$ is the specific emissivity, $\chi_\nu$ is the specific
extinction, and $c$ is the speed of light. In the context of atomic resonance
line scattering,
$\eta_\nu=(c/4\pi)(\chi_\nu/\phi_\nu)\int\,d\nu^\prime\,R(\nu^\prime,\nu)u_{\nu^\prime}$,
where $u_\nu$ is the specific energy density of the radiation field
and $R(\nu^\prime,\nu)$ is the redistribution function for scattering
a photon of frequency $\nu^\prime$ to $\nu$ in the rest frame of the
atom. The effect of atomic recoils is incorporated into $R(\nu^\prime,\nu)$.

In a cosmological context, the sources are considered isotropic and homogeneously distributed, resulting in a homogeneous and isotropic radiation field averaged over sufficiently large scales. The radiative transfer equation may then be integrated over angle, giving
\begin{equation}
\frac{D u_\nu}{D t}
=\frac{\chi_0c}{h\nu_0}h\nu
\biggl[\int_0^\infty d\nu' R(\nu',\nu)u_{\nu'}
\phantom{\biggl[}-\varphi(\nu) u_\nu\biggr]
+ h\nu S(\nu),
\label{eq:RThom}
\end{equation}
where $\chi_0$ is $\chi_\nu$ evaluated at the resonance line center frequency $\nu_0$, $\varphi(\nu)$ is the absorption line profile and a source $S_\nu$ is included in the emissivity. In the diffusion approximation, the equation is approximated by expanding the scattering integrand producing a diffusion partial differential equation for the radiation field. As emphasized by \cite{2006ApJ...647..709R}, the expansion is not unique, with alternative definitions for the diffusion coefficient. While the solutions agree far from line center, it is not possible to conserve both photon number and energy when integrating over all frequencies in the diffusion approximation. An alternative approximation is to use the Fokker-Planck approximation which does conserve both photon number and energy, but second order solutions yield unphysical artifacts within the core, although these may be mitigated while retaining the energy exchange term with the gas \citep{2006MNRAS.370.2025M}.
%Extending to higher order in a Kramers-Moyal expansion fails because of the Pawula theorem, which requires all orders to ensure non-negative scattering rates if carried above second order (REF:Riskin ...).

Integrating Eq.~(\ref{eq:RTgen}) over angle and frequency, it follows
that the exact rate of energy transfer per volume between the
radiation field and the gas is given by

\begin{equation} G=P_l n_l
\frac{h\nu_0}{m_a c^2} h\nu_0 \left(1-\frac{T}{T_L}\right),
\label{eq:GheatEx}
\end{equation}
\citep{2006MNRAS.370.2025M}, where $P_l$ is the total photon
scattering rate per atom in the lower state with number density $n_l$,
and $T_L$ is the light temperature governing the hydrogen spin
temperature through the Wouthuysen-Field effect
\citep{1952AJ.....57R..31W, 1958PROCIRE.46..240F}. (In
\cite{2006MNRAS.370.2025M}, Eq.[\ref{eq:GheatEx}] is expressed using the
mean harmonic color temperature $\langle T_u\rangle_H$, but there it
is shown $T_L=\langle T_u\rangle_H$ to high precision.) The factor
$1-T/T_L$ governs the efficiency of Lyman-$\alpha$ photon recoil heating,
including the possibility of cooling when $T_L<T$. When a typical continuum source first turns on, $T_L\gg T$ and the heating rate is large during Cosmic Dawn \citep{1997ApJ...475..429M}. In a static medium,
$T_L\rightarrow T$ after 100-1000 scattering times and heating shuts off
\citep{2006MNRAS.370.2025M}.

In an expanding medium, $T_L$ freezes out before reaching $T$
\citep{2004ApJ...602....1C, 2006MNRAS.370.2025M}. In the diffusion
approximation of \cite{1994ApJ...427..603R} and following \cite{2004ApJ...602....1C} and \cite{2006MNRAS.372.1093F}, the efficiency may be re-expressed as

\begin{equation}
\left(1-\frac{T}{T_L}\right) \approx \frac{\gamma}{\epsilon}\frac{I_c}{S_\alpha},
\label{eq:effDA}
\end{equation}
where $\gamma=\nu_0H/\chi_0c$ is the Sobolev parameter for Hubble parameter
$H$ and $\epsilon=h\nu_0/(2k_\mathrm{B}Tm_ac^2)^{1/2}$ is the recoil
parameter for an atomic mass $m_a$. Here $I_c =
\int\,dx\,[1-u(x)/u_\infty]$ where $u(x)=u_\nu d\nu/dx$ for
$x=(\nu-\nu_0)/\Delta\nu_D$ with Doppler width
$\Delta\nu_D=(2k_\mathrm{B}T/m_a)^{1/2}\nu_0/c$, $u_\infty$ is the
energy density in the background radiation field far from line center,
and $S_\alpha$ is a small correction to $u(x)$ across the line center arising from atomic recoils.

\section{Application}
\label{sec:application}

\begin{figure}
\scalebox{0.5}{\includegraphics{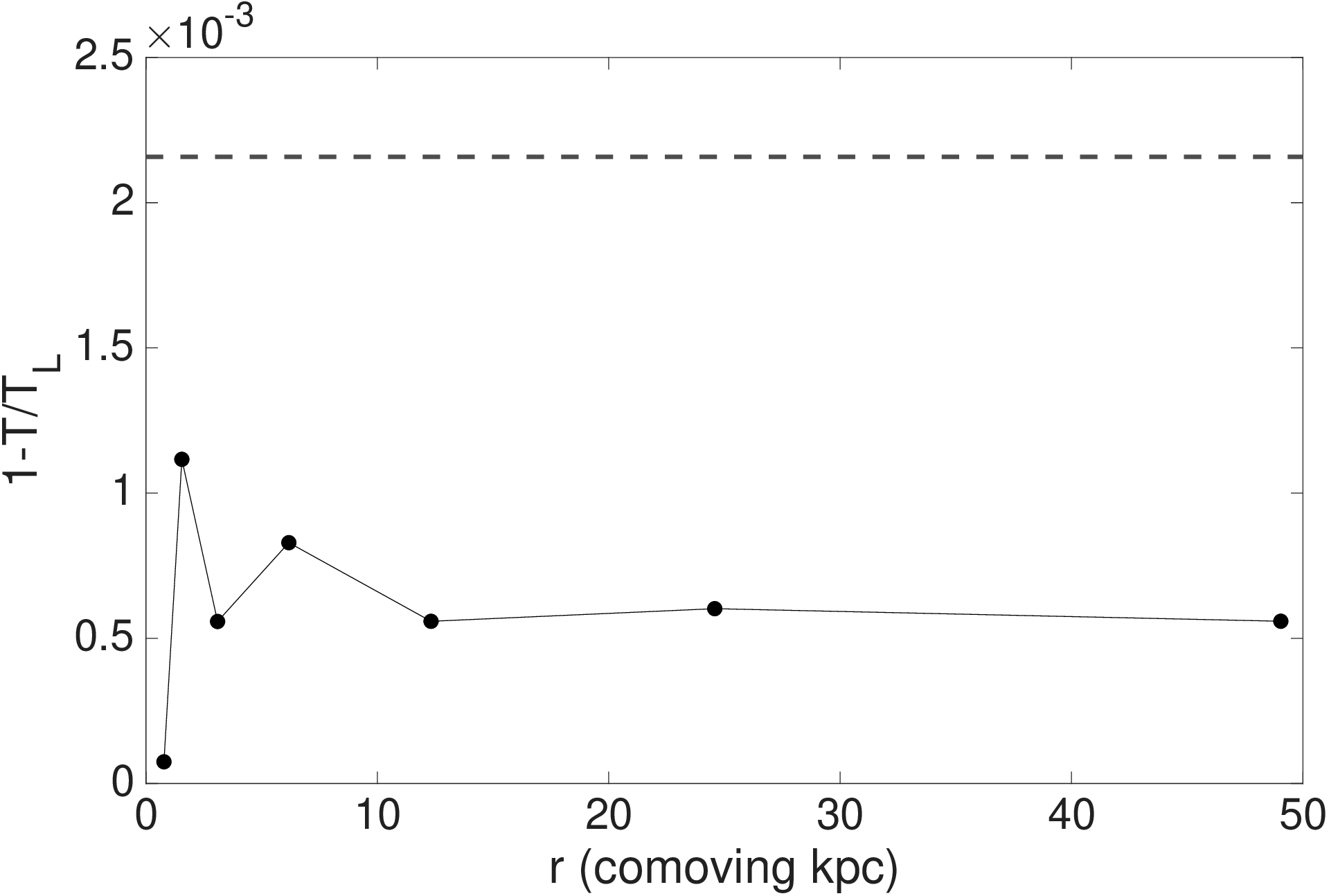}}
\caption{The radial dependence of the Lyman-$\alpha$ photon recoil heating efficiency for a point continuum source at $z=10$ in intergalactic gas with temperature $T=10$~K. The points show the results solving the full integro-differential radiative transfer equation. The efficiency approaches a constant value distant from the source, corresponding to the efficiency expected for a homogeneous distribution of such sources. The dashed line shows the estimate using the diffusion approximation to the radiative transfer equation for homogeneously distributed sources.
}
\label{fig:efficiency}
\end{figure}

\cite{2012MNRAS.426.2380H} solve the steady-state form of
Eq.~(\ref{eq:RTgen}) in spherical symmetry in full integro-differential
form numerically for a point continuum source at a typical Cosmic Dawn
redshift of $z=10$ and Intergalactic Medium temperature $T=10$~K, for which $\gamma
\simeq 1.27\times10^{-6}$ and $\epsilon\simeq0.0080$. The resulting
energy density profile is shown in their fig.~10 and tabulated in
their table~D6. The corresponding light temperature
$T_L\simeq10.006$~K, nearly constant with distance from the source, as
shown in Fig.~\ref{fig:efficiency}. Since the radiative transfer equation is linear in
the intensity and source term, the radiation field from a homogeneous
distribution of such sources will also have $T_L\simeq10.006$~K. The
Lyman-$\alpha$ recoil heating efficiency factor for a uniform distribution of
such sources is then $1 - T/T_L \simeq 0.0006$.

For comparison, using $1-T/T_L=(\gamma/\epsilon)I_c/S_\alpha$, the
solution of \cite{2006MNRAS.372.1093F} to the radiative transfer
equation in the diffusion approximation gives $1-T/T_L\simeq0.002$, a
factor 3.8 too large. (The light temperature is given as
$T_L=10.02$~K.) It is currently an open question whether a different
approximation to the full radiative transfer equation would yield more
accurate recoil heating estimates for including in Cosmic Dawn
numerical models.

%%%%%%%%%%%%%%%%%%%%%%%%%%%%%%%%%
%\bigskip  
%\section*{acknowledgments}

%The author thanks S. Mittal and  G. Kulkarni for exchanges that led
%the author to revisit the subject. The author also acknowledges
%support from the UK Science and Technology Facilities Council, Consolidate Grant ST/R000972/1.

%%%%%%%%%%%%%%%%%%%%%%%%%%%%%%%%%

\bibliographystyle{aasjournal}
\bibliography{ms}

%\label{lastpage}

\end{document}